\documentclass[letterpaper, 10 pt, conference]{ieeeconf}  

\IEEEoverridecommandlockouts                              

\usepackage{graphics} 
\usepackage{epsfig} 
\usepackage{mathptmx} 
\usepackage{times} 
\usepackage{amsmath} 
\usepackage{amssymb}  
\usepackage{subfigure}
\usepackage[ruled,linesnumbered]{algorithm2e}
\usepackage{multirow}
\usepackage{color}
\usepackage{float} 

\DeclareMathAlphabet{\mathcal}{OMS}{cmsy}{m}{n}

\title{\LARGE \bf
 A Distributed Clustering Algorithm based on Coalition Game for Intelligent Vehicles}

\author{Weiyi Yang, Xiaolu Liu, Lei He$^*$, Yonghao Du, Yingwu Chen
\thanks{*This research was supported by the National Natural Science Foundation of China (72001212), the Young Elite Scientists Sponsorship Program by CAST (2022QNRC001)}
\thanks{Weiyi Yang, Xiaolu Liu, $^{*}$Lei He, Yonghao Du, and Yingwu Chen are with the College of Systems Engineering, National University of Defense Technology, Changsha 410073, Hunan Province, China. 
{\tt\small (yangweiyi15, lxl\_sunny, helei, duyonghao15, ywchen) @nudt.edu.cn}
}}

\begin{document}

\maketitle
\thispagestyle{empty}
\pagestyle{empty}

\begin{abstract}

In the context of Vehicular ad-hoc networks (VANETs), the hierarchical management of intelligent vehicles, based on clustering methods, represents a well-established solution for effectively addressing scalability and reliability issues. The previous studies have primarily focused on centralized clustering problems with a single objective.  However, this paper investigates the distributed clustering problem that simultaneously optimizes two objectives: the cooperative capacity and management overhead of cluster formation, under dynamic network conditions. Specifically, the clustering problem is formulated within a coalition formation game framework to achieve both low computational complexity and automated decision-making in cluster formation. Additionally, we propose a distributed clustering algorithm (DCA) that incorporates three innovative operations for forming/breaking coalition, facilitating collaborative decision-making among individual intelligent vehicles. The convergence of the DCA is proven to result in a Nash stable partition, and extensive simulations demonstrate its superior performance compared to existing state-of-the-art approaches for coalition formation.

\end{abstract}


\section{INTRODUCTION}

The increasing integration of intelligent vehicles with high-performance sensors, processing and communication devices, along with constant Internet connectivity, has facilitated the emergence of the Internet of Vehicles (IoV)\cite{Han.2023,Chattopadhyay.2023}. With advanced IoV technologies, vehicles can efficiently exchange data\cite{Han.2023} and collaboratively sense\cite{Wang.2023c,Zhao.2023,Zhu.2022,Wu.2024}, communicate\cite{Falahatraftar.2023,Liu.2024}, compute\cite{Lang.2022,Lang.2023,Li.2023,Liu.2023}, and control\cite{Xiao.2022}. This further propels technologies for transportation 5.0 such as autonomous driving\cite{Chattopadhyay.2023,Yang.2023} and smart infrastructure services\cite{Han.2024}. To manage a large number of intelligent vehicles, Vehicular ad-hoc networks (VANETs) has garnered significant attention from both academia and industry.

The VANETs typically employ wireless technology for facilitating Vehicle-to-Vehicle (V2V) or Vehicle-to-Infrastructure (V2I) communications. However, the existing organizational and operational structures of VANETs face significant challenges, including the highly dynamic topology, concerns regarding security and privacy, as well as diverse quality of service (QoS) requirements. These challenges become particularly pronounced when a centralized management system is utilized. To deal with these challenges, the clustering of adjacent vehicles into groups presents significant potential for enhancing scalability, coordination, and maintainability in VANETs \cite{ref1}.

The vehicles in VANETs can be categorized into different clusters, with each cluster consisting of at least one designated cluster head (CH) responsible for facilitating inter-cluster connectivity, as well as multiple cluster members (CMs) that collaborate with other CMs within the same cluster to support a variety of cooperative services (e.g., cooperative communication\cite{Das.2016}, spectrum allocation or popular content distribution service\cite{Wang.2020}). In the cluster formation, the link capacity between vehicles could directly influence the communication performance among vehicles, thereby impacting the quality of service (QoS) for the entire VANET.  Therefore, when clustering VANETs, it is crucial to consider the \textbf{cooperative capacity} of cluster formation based on link capacity as an optimization objective.
Besides, the adaptive clustering process under dynamic network conditions inevitably leads to significant \textbf{management overhead}, particularly in terms of communication power consumption. Therefore, it is also crucial to optimize this aspect, especially for large-scale VANETs with limited resources available for each vehicle.

Motivated by the aforementioned backgrounds, this paper investigates the VANET clustering problem that jointly optimizes cooperative capacity and management overhead based on dynamic network conditions. The challenges for VANET clustering problem are threefold. 1) In contrast to previous studies that focus solely on power consumption, a comprehensive multi-objective clustering approach is necessary to achieve a balance between maximizing cluster cooperative capacity and minimizing management overhead. 2) The large-scale nature of VANET introduces significant computational complexity for centralized optimization clustering approaches, necessitating the deployment of distributed clustering approaches on individual intelligent vehicles. However, ensuring convergence and solution optimality in such distributed algorithms is often challenging. 3) The high level of dynamism in intelligent vehicles introduces significant uncertainty to the VANET clustering problem, necessitating an adaptive approach capable of handling dynamic changes in the VANET environment.

In response to the above challenges, we propose an adaptive game-theory-based clustering algorithm for VANET. The main contributions of this paper are threefold. 1) We formulate the VANET clustering problem with two novel objectives (i.e., the cluster cooperative capacity and he management overhead). 2) The VANET clustering problem is reexamined within the framework of coalition games, and a fully distributed clustering algorithm (DCA) is proposed to be implemented at the individual vehicle level. Furthermore, the proof of its convergence to Nash stability is subsequently provided. 3) The performance superiority of DCA is investigated through extensive simulations, in comparison with four baseline approaches.

\section{Related works}
The clustering problems in VANETs have been extensively investigated over the past few years. Most existing works in VANET clustering have focused on the offline centralized algorithm wherein a central controller aggregates global information for obtaining the final clusters. The typical offline centralized approach involves a two-stage optimization process: firstly, selecting cluster heads based on various weighted indicators that prioritize nodes (vehicles) with high weights, and then assigning the remaining nodes to join these cluster heads. Various heuristic algorithms have been employed in pervious works to optimize the aforementioned two-stage process. For instance, Ref \cite{Khan.2019} proposed a nature-inspired Moth Flame algorithm to optimize the formation of vehicle clusters.

The lack of robustness and scalability makes centralized approaches unsatisfied for addressing the clustering problem in VANETs involving nodes in a dynamic mobility environment \cite{ref2}. Thus, the establishment of a distributed self-organized approach is expected , wherein a node accomplishes the global objective through interaction and communication with other nodes \cite{ref3}. However, the assistance of a centralized controller is indispensable for ensuring a high-quality solution. Consequently, the primary challenge in a distributed approach lies in devising an individual decision-making algorithm that aligns with the global objective on a theoretical level.

Due to the dynamic nature of VANET, centralized optimization would result in high computational complexity and significant control costs. To tackle this challenge, distributed and autonomous systems are required. Therefore, coalition game theory is employed to investigate player behavior during cooperation and provide a relevant framework for ad-hoc network clustering. The coalitional formation game framework is a powerful tool to study the behavior of players when they cooperate, and provide a relevant framework for ad-hoc network clustering. This framework has also been proposed for several applications such as: satellite swarm \cite{ref6} or vehicle-to-vehicle communications \cite{ref7} and massive machine-to-machine networks \cite{ref8}. Note that, it has to be mentioned that the impact of management overhead is not taken into account in most previous works.

To summarize, despite the extensive research on clustering problems in VANETs, most existing studies primarily focus on offline centralized approaches and isolated single-objective optimization. Therefore, this study proposes a distributed approach based on coalition game theory to address the multi-objective clustering problem in VANET, taking into account the dynamic nature of the environment.

\section{Problem description and modeling}
\subsection{System model}
The system model of the VANET is shown in Fig. 1, which involves the typical entities such as intelligent vehicles (IVs) and roadside units (RSUs). The communication range of RSU 1 is depicted by a yellow irregular elliptical shape, while that of IV 1 is represented by an orange irregular elliptical shape. In cases where the sender entity is beyond communication range, the message may be received through multi-hop cooperative entities by either the IV or RSU. Each entity possesses a distinct communication range, with RSUs typically exhibiting a larger communication range.
\begin{figure}[!htb]
	\centering
	\includegraphics[width=0.5\textwidth]{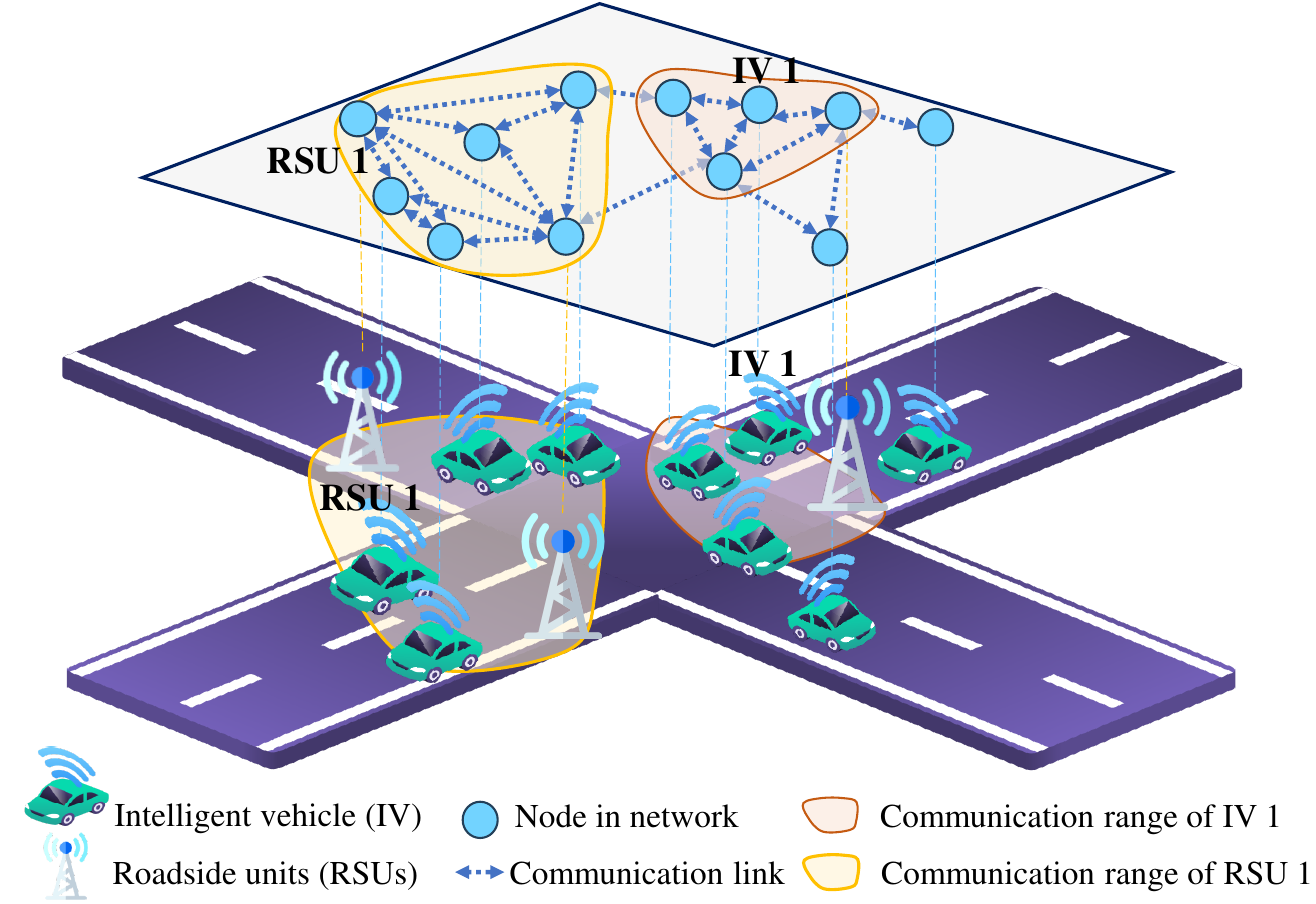}
	\caption{Diagram of the system model in VANET}
	\label{f0}
\end{figure}

It is assumed that the links in VANET only exist between two entities with line-of-sight (LOS) connections, and the path loss expressions for the link between entity $i$ and entity $j$, separated by a distance of $d_{i,j}$, are given as

\begin{equation}
    PL({{d}_{i,j}})={{({}^{\lambda }/{}_{4\pi })}^{{{\alpha }_{L}}}}\cdot {{\left( {{d}_{i,j}} \right)}^{-{{\alpha }_{L}}}}
\end{equation}

, where ${{({}^{\lambda }/{}_{4\pi })}^{{{\alpha }_{L}}}}$ is constant that represents the path loss of unit length of LOS links, and $\lambda $ is the wavelength. $\alpha_{L} $ is the path loss of LOS which is affected by the environment scenario. For the link ($i$, $j$), given the link transmit power $P_t$, the maximum directive antenna gain $G_0$ and the channel power gain $h_{i,j}$ of link ($i$, $j$), the SINR at receiver of link ($i$, $j$) is given as follows
\begin{equation}
    {{\Lambda }_{i,j}}=\frac{{{P}_{t}}{{G}_{0}}{{h}_{i,j}}\cdot PL({{d}_{i,j}})}{{{N}_{0}}+{{I}_{i,j}}}
\end{equation}

, where ${{N}_{0}}$ is assumed as additive white Gaussian noise and $I_{i,j}$ is the sum interference of other concurrent links:
\begin{equation}
    {{I}_{i,j}}={{P}_{t}}{{G}_{0}}\cdot {{({}^{\lambda }/{}_{4\pi })}^{{{\alpha }_{L}}}}\sum\limits_{k\in N\backslash \{i\}}{{{h}_{k,j}}\cdot {{\left( {{d}_{k,j}} \right)}^{-{{\alpha }_{L}}}}}
\end{equation}
Therefore, the link capacity $Rt_{i,j}$ of link $(i, j)$ between entity $i$ and $j$ is given by
\begin{equation}
    R{{t}_{i,j}}=W{{\log }_{2}}\left( 1+{{\Lambda }_{i,j}} \right)
\end{equation}
\subsection{Problem description}
The vehicular ad hoc networks (VANET) $\mathcal{W}$ considered in this paper is defined by the set of nodes $\mathsf{\mathcal{N}}$ and the set of edges $\mathsf{\mathcal{E}}$, where both the RSUs and IVs are considered as nodes. Two nodes are neighbors if they are in the communication range of each other, then $\left( i,j \right)\in \mathsf{\mathcal{E}}$. A clustering solution leads to a partition $\Pi =\left\{ {\mathcal{S}_{1}},{\mathcal{S}_{2}},...,{\mathcal{S}_{m}},...,{\mathcal{S}_{M}} \right\}$, where $M$ represents the number of clusters. In each cluster, ${\mathcal{S}_{m}}\in \Pi$ exist only one cluster head and several cluster members, and the set of cluster heads is ${{\mathsf{\mathcal{N}}}_{head}}$. 

The clustering architecture is illustrated in Fig. 2. Generally, the organization of nodes in a clustered network can be divided into two layers: inter-cluster cooperation and intra-cluster cooperation. Inter-cluster cooperation involves communication and interaction at the cluster head level, where all information within a cluster is consolidated \cite{ref9}.On the other hand, intra-cluster cooperation refers to the interaction among nodes within a cluster. In this paper, the node in the cluster that responds to communicate with other clusters and engages in inter-cluster cooperation is referred to as the 'cluster head' node.

\begin{figure}[!htb]
	\centering
	\includegraphics[width=0.48\textwidth]{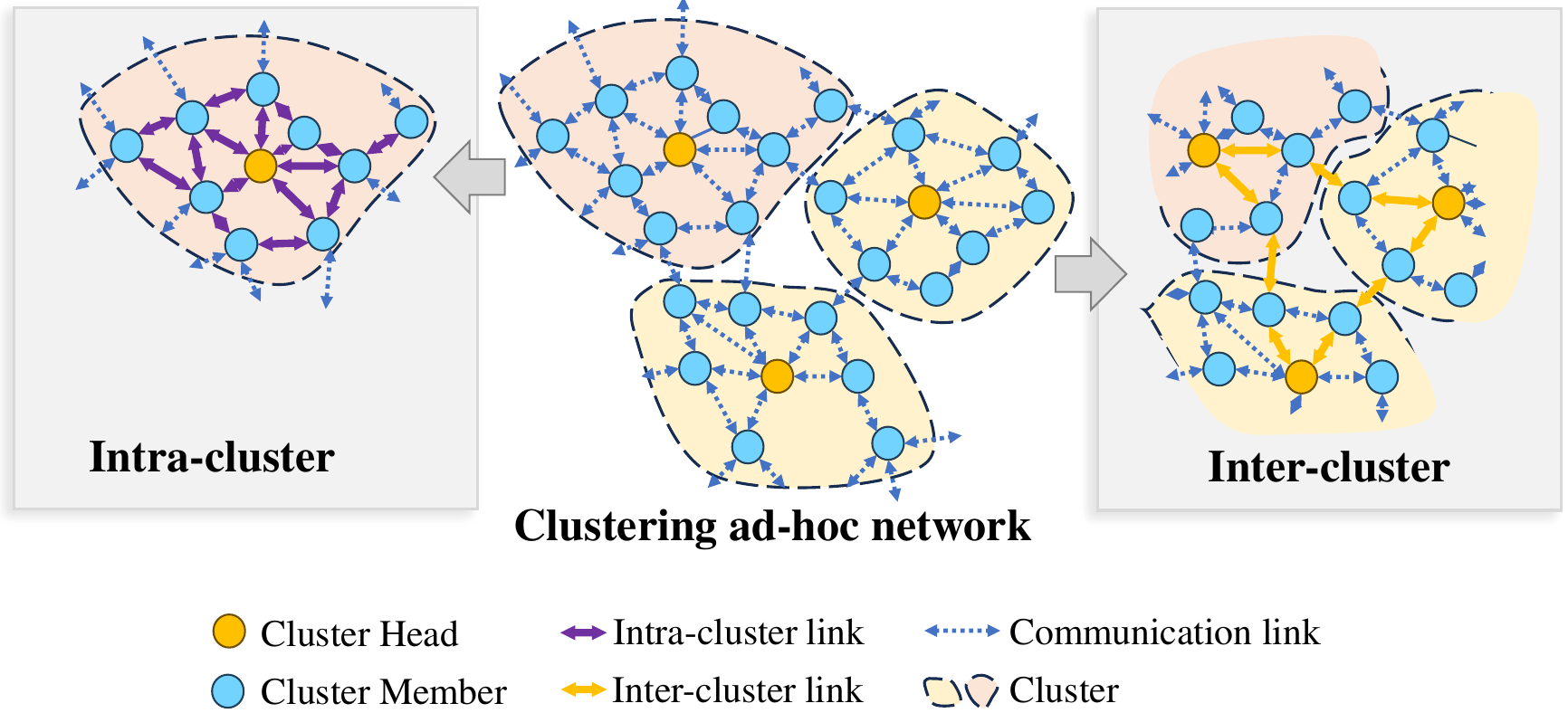}
	\caption{ Clustering architecture diagram.}
	\label{f1}
\end{figure}

\subsection{Problem Modeling}
In this section, two novel key performance indicators are introduced to evaluate the performance of a cluster scheme. Subsequently, the VANET clustering problem is formulated as a constraint optimization problem, where constraints are imposed on both the number of nodes and diameter for each cluster. 

The cluster capacity is defined as the sum of link capacity \cite{ref2}, and the link capacity is further expressed as the minimum achieve communication rate for the shortest path between two nodes. For instance, ${{q}_{i,j}}$ is the shortest path between node $i$ and node $j$, which can be expressed as ${{q}_{i,j}}=\left( \left( i,{{i}_{1}} \right),\left( {{i}_{1}},{{i}_{2}} \right),...,\left( {{i}_{h}},j \right) \right)$. The link capacity between node $i$ and node $j$ is defined as 

\begin{equation}
\begin{array}{l}
{{\kappa }_{ij}}=\frac{1}{\alpha \cdot \left| {{q}_{i,j}} \right|}\min \left( R{{t}_{i,{{i}_{1}}}},R{{t}_{{{i}_{1},}{{i}_{2}}}},...,R{{t}_{{{i}_{h}},j}} \right)
\end{array}
\label{eq1}  
\end{equation}

, where $R{{t}_{i,j}}$ is the achieve communication rate for the link between node $i$ and node $j$, which is calculated as (4), and $\alpha$ is a multi-hop path loss parameter which satisfied $\alpha >1$. 

The total cluster cooperative capacity ${{\chi }_{total}}$ includes the inter-cluster cooperative capacity ${{\chi }_{inter}}$ and the intra-cluster cooperative capacity ${{\chi }_{intra}}$
\begin{equation}
    {{\chi }_{total}}=\sum\limits_{{\mathcal{S}_{m}}\in \Pi }{{{\chi }_{inter}}(m)}+\sum\limits_{{\mathcal{S}_{m}}\in \Pi }{{{\chi }_{intra}}(m)}
\end{equation}
\begin{equation}
    {{\chi }_{inter}}(m)=\beta \sum\limits_{j\in {{\mathsf{\mathcal{N}}}_{head}}}{{{\kappa }_{ij}}}\quad i_{m}^{h}\in {{\mathsf{\mathcal{N}}}_{head}}\cap {{\mathsf{\mathcal{S}}}_{m}}
\end{equation}
\begin{equation}
    {{\chi }_{intra}}(m)=\sum\limits_{i\in {{\mathsf{\mathcal{S}}}_{m}}}{\sum\limits_{j\in {{\mathsf{\mathcal{S}}}_{m}}\cap {{\mathsf{\mathcal{R}}}_{i}}}{{{\kappa }_{ij}}}}
\end{equation}

, where ${{\chi }_{inter}}(m)$ is inter-cluster cooperative capacity reflected by the link capacity of all cluster heads, $i_{m}^{h}$ refers to the head node of cluster ${{\mathsf{\mathcal{S}}}_{m}}$, $\beta $ represent the parameter of inter-cluster link capacity, ${{\chi }_{intra}}(m)$ is the intra-cluster cooperative capacity for cluster ${{\mathsf{\mathcal{S}}}_{m}}$, which is the sum of link capacities between each node in ${{\mathsf{\mathcal{S}}}_{m}}$.

For instance, as depicted in Fig. 3, there exist three clusters with nodes 1, 4, and 6 serving as cluster heads respectively. The inter-cluster communication capacity among them can be expressed as $\beta \left( {{\kappa }_{1,4}}+{{\kappa }_{4,6}}+{{\kappa }_{1,6}}+{{\kappa }_{4,1}}+{{\kappa }_{6,4}}+{{\kappa }_{6,1}} \right)$. The intra-cluster cooperative capacity among them can be expressed as $R{{t}_{3,1}}+R{{t}_{2,4}}+R{{t}_{3,4}}+R{{t}_{3,5}}+R{{t}_{4,6}}+R{{t}_{1,2}}+R{{t}_{1,4}}+R{{t}_{5,6}}$.

\begin{figure}[htbp]
\centering
\setlength{\abovecaptionskip}{0.cm}
\subfigure[inter-cluster]{
\includegraphics[width=3cm]{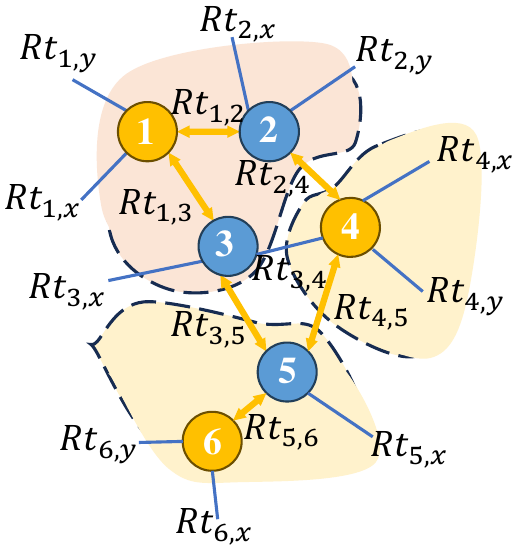}
\label{Fig1a}
}
\quad
\subfigure[intra-cluster]{
\includegraphics[width=3.5cm]{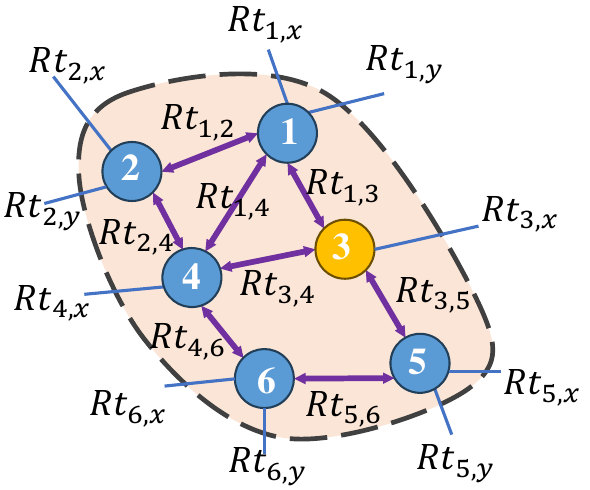}
\label{Fig1c}
}
\caption{Diagram of an example of cooperative capacity calculation}\label{Fig1}
\end{figure}

Same as Ref. \cite{ref6}, we employ the routing overhead generated by formulating and maintaining clusters as the clustering management overhead. The cluster management overhead is the sum of inter-cluster management overhead ${{E}_{inter}}$ and intra-cluster management overhead ${{E}_{intra}}$ as well
\begin{equation}
    {{E}_{total}}=\sum\limits_{{\mathcal{S}_{m}}\in \Pi }{{{E}_{inter}}(m)}+\sum\limits_{{\mathcal{S}_{m}}\in \Pi }{{{E}_{intra}}(m)}
\end{equation}
\begin{equation}
    {{E}_{inter}}(m)=\sum\limits_{i\in {{\mathsf{\mathcal{N}}}_{head}}}{\sum\limits_{j\in {{\mathsf{\mathcal{N}}}_{head}}/\left\{ i \right\}}{{{V}_{inter}}}}={{V}_{inter}}\cdot \left| {{\mathsf{\mathcal{N}}}_{head}} \right|\cdot \left( \left| {{\mathsf{\mathcal{N}}}_{head}} \right|-1 \right)
\end{equation}
\begin{equation}
    {{E}_{intra}}(m)=\sum\limits_{i\in {{\mathsf{\mathcal{S}}}_{m}}}{\sum\limits_{j\in {{\mathsf{\mathcal{S}}}_{m}}/\{n\}}{{{V}_{intra}}}}={{V}_{intra}}\cdot \left| {{\mathsf{\mathcal{S}}}_{m}} \right|\cdot \left( \left| {{\mathsf{\mathcal{S}}}_{m}} \right|-1 \right)
\end{equation}

, where ${{E}_{inter}}(m)$ and ${{E}_{intra}}(m)$ refers to the management overhead of cluster ${{\mathsf{\mathcal{S}}}_{m}}$ for inter-cluster and intra-cluster interaction and maintaining. ${{V}_{inter}}$ and ${{V}_{intra}}$ refer to the data volume for inter-cluster and intra-cluster management overhead.

Considering the delay of intra-cluster and inter-cluster communication, the cluster size $\left| {\mathcal{S}_{m}} \right|$ and diameter $D({\mathcal{S}_{m}})$ need to be constrained for each cluster ${\mathcal{S}_{m}}\in \Pi $. Therefore, the optimization problem P1 for VANET clustering problem can be formulated as follows.

\begin{equation}
\max \ {{G}_{1}}=\left( 1-\zeta  \right)\overline{{{\chi }_{total}}}+\zeta \cdot \overline{{{E}_{totol}}}
\end{equation}
\begin{equation}
\left| {\mathcal{S}_{m}} \right|\le {{N}_{\max }}\quad \forall {\mathcal{S}_{m}}\in \Pi 
\end{equation}
\begin{equation}
D({\mathcal{S}_{m}})\le {{D}_{\max }}\quad \forall {\mathcal{S}_{m}}\in \Pi 
\end{equation}
\begin{equation}
\overline{{{\chi }_{total}}}=\frac{{{\chi }_{total}}}{{{\chi }_{\max }}},\quad \overline{{{E}_{totol}}}=\frac{{{E}_{\max }}-{{E}_{total}}}{{{E}_{\max }}}
\end{equation}
, where the weigh coefficient $\zeta \in [0,1]$ is used to make a trade-off between the two indicators. Besides, ${{N}_{\max }}$ and ${{D}_{\max }}$ are the predefined maximum number of cluster members and maximum cluster diameter respectively. $\overline{{{\chi }_{total}}}$ and $\overline{{{E}_{totol}}}$ represent the normalized form of ${{\chi }_{total}}$ and ${{E}_{totol}}$.

\section{Distributed Coalition Agorithm}
The coalition formation game (CFG) framework is adopted to reformulate the optimization problem P1. Then, a distributed clustering algorithm is proposed to solve it. The convergence of the algorithm is subsequently proved. 
\subsection{Coalition formation game}
CFG involves a set of players who want to cooperate by forming coalitions in order to improve their positions in the game. First, the clustering problem can be reformulated as follows:

\begin{equation}
\mathsf{\mathcal{G}}=\left( \mathsf{\mathcal{N}},{{\left\{ {{a}_{n}} \right\}}_{n\in \mathsf{\mathcal{N}}}},V \right)
\end{equation}

, where $\mathsf{\mathcal{N}}=\left\{ 1,2,...,N \right\}$ represent the players (nodes) set. ${{\left\{ {{a}_{i}} \right\}}_{i\in \mathsf{\mathcal{N}}}}={\mathcal{S}_{i}}\otimes {{h}_{i}}$ represents the action set of players, where $\otimes $ is the Cartesian product. ${\mathcal{S}_{i}}\in \left\{ {\mathcal{S}_{m}} \right\}$ represents the available coalition (i.e., cluster) that player $i$ could join. ${{h}_{i}}$ refers to the indicator actions: ${{h}_{i}}=1$ indicates that node $i$ is the head node whereas ${{h}_{i}}=0$ otherwise; Define ${{a}_{-i}}\in {{a}_{1}}\otimes {{a}_{2}}\otimes \ldots \otimes {{a}_{i-1}}\otimes {{a}_{i+1}}\ldots \otimes {{a}_{N}}$ as collections of strategies for all nodes except node $i$.

$v({\mathcal{S}_{m}})$ is the coalition function, indicating the total-payoff generated by any ${\mathcal{S}_{m}}$ and is defined as
\begin{equation}
    v({{\mathsf{\mathcal{S}}}_{m}})=\left\{ \begin{matrix}
   \left( 1-\zeta  \right)\frac{{{\chi }_{total}}(m)}{{{\chi }_{\max }}}+\zeta \cdot \frac{\frac{{{E}_{\max }}}{M}-{{E}_{total}}(m)}{{{E}_{\max }}}\ if\;(13),(14)\;hold  \\
   0\quad \quad \quad \quad \quad \quad \quad \quad \quad \quad \quad \quad \quad \quad \quad \quad \quad otherwise  \\
\end{matrix} \right.
\end{equation}
Given a feasible coalition partition $\Pi :=\left\{ {\mathcal{S}_{1}},...,{\mathcal{S}_{M}} \right\}$, the sum of the coalition value function is express as 

\begin{equation}
\begin{aligned}
  & \sum\limits_{{\mathcal{S}_{m}}\in \Pi }{v({\mathcal{S}_{m}})}=\sum\limits_{{\mathcal{S}_{m}}\in \Pi }{\left( 1-\zeta  \right)\frac{{{\chi }_{total}}(m)}{{{\chi }_{\max }}}+\zeta \cdot \frac{{{{E}_{\max }}}/{M}\;-{{E}_{total}}(m)}{{{E}_{\max }}}} \\ 
 & =\left( 1-\zeta  \right)\frac{\sum\limits_{{\mathcal{S}_{m}}\in \Pi }{{{\chi }_{total}}(m)}}{{{\chi }_{\max }}}+\zeta \cdot \frac{{{E}_{\max }}-\sum\limits_{{\mathcal{S}_{m}}\in \Pi }{{{E}_{total}}(m)}}{{{E}_{\max }}} \\ 
 & =\left( 1-\zeta  \right)\overline{{{\chi }_{total}}}+\zeta \cdot \overline{{{E}_{totol}}}\quad \\
 & \forall {\mathcal{S}_{m}}\in \Pi ,\left| {\mathcal{S}_{m}} \right|\le {{N}_{\max }},D({\mathcal{S}_{m}})\le {{D}_{\max }} \\ 
\end{aligned}
\end{equation}
, which conforms to the model of problem P1.

In this regard, the rules governing coalition formation and dissolution among nodes are defined. However, these traditional operations, such as merge operation and split operations, pose challenges during later stages of coalition adjustment, leading to potential issues such as partial traps or invalid calculations. To address these problems associated with traditional operations, three novel operations set ${{P}_{i}}$ of node $i$ are designed for coalition formation as follows:

\textbf{Definition 1.} \textit{ For any player $i,i_\mathcal{S}^{h}\in \mathcal{S}$, a head node election operation ${{\varsigma }_\mathcal{S}}(i,i_\mathcal{S}^{h})\in {{P}_{i}}$ is defined is the player $i$ replace the original head node $i_\mathcal{S}^{h}$ becomes the  head node of the cluster $\mathcal{S}$}

\begin{equation}
{{\varsigma }_\mathcal{S}}(i,i_\mathcal{S}^{h}):{{\mathsf{\mathcal{N}}}_{head}}\to \left\{ {{\mathsf{\mathcal{N}}}_{head}}\cup \left\{ i \right\} \right\}/\left\{ i_\mathcal{S}^{h} \right\}
\end{equation}

\textbf{Definition 2.} \textit{ For any players $i\in \mathcal{S}$ and ${i}'\in {\mathcal{S}'}$ and all coalition $\mathcal{S}$ and ${\mathcal{S}'}$, a bidirectional switch operation ${{\sigma }_{\mathcal{S},{\mathcal{S}'}}}\left( i,{i}' \right)\in {{P}_{i}}$ is defined if every players in the newly formed coalition  $\mathcal{S}$ and ${\mathcal{S}'}$}

\begin{equation}
{{\sigma }_{\mathcal{S},{\mathcal{S}'}}}\left( i,{i}' \right):{i}'\to \mathcal{S}\backslash \left\{ i \right\},i\to {\mathcal{S}'}\backslash \left\{ {{i}'} \right\}
\end{equation}

\textbf{Definition 3} \textit{ For the player $i\in \mathcal{S}$ and $i'\in {\mathcal{S}'}$ and all coalition $\mathcal{S}$ and ${\mathcal{S}'}$, a replace operation ${{\rho }_{\mathcal{S},{\mathcal{S}'}}}\left( i,i' \right)\in {{P}_{i}}$  is defined  if every user in the newly formed coalition $\mathcal{S}$ and ${\mathcal{S}'}$}

\begin{equation}
{{\rho }_{\mathcal{S},{\mathcal{S}'}}}\left( i,i' \right):i\to {\mathcal{S}'}\backslash \left\{ i' \right\},i'\to \left\{ i' \right\}
\end{equation}

The above two operations also include directional replacement operation and coalition joining operation by replacing the coalition ${\mathcal{S}'}$ and node ${i}'$ with $\varnothing $. For instance, ${{\sigma }_{\mathcal{S},{\mathcal{S}'}}}\left( i,\varnothing  \right)$ refers to the coalition $\mathcal{S}$ and ${\mathcal{S}'}$ switch only node ${i}$.

\textbf{Definition 4} \textit{ A operation ${{\mathsf{\mathcal{P}}}_{i,\mathcal{S}}}\in {{P}_{i}}$, no matter what type of operation it is, happens when there is an operation gain $g\left( {{\mathsf{\mathcal{P}}}_{i,\mathcal{S}}} \right)>0$}
\begin{equation}
g\left( {{\varsigma }_\mathcal{S}}(i,i_\mathcal{S}^{h}) \right):=v\left( \mathcal{S}\left| {{h}_{i}}=1 \right. \right)-v\left( \mathcal{S}\left| {{h}_{i_\mathcal{S}^{h}}}=1 \right. \right) 
\end{equation}
\begin{multline}
g\left( {{\sigma }_{\mathcal{S},{\mathcal{S}'}}}\left( i,{i}' \right) \right):=v\left( \left( \mathcal{S}\backslash \left\{ i \right\} \right)\cap \left\{ i' \right\} \right)+ \\
v\left( \left( {\mathcal{S}'}\backslash \left\{ i' \right\} \right)\cap \left\{ i \right\} \right)-v\left( \mathcal{S} \right)-v\left( {{\mathcal{S}'}} \right) 
\end{multline}
\begin{multline}
    g\left( {{\rho }_{\mathcal{S},{\mathcal{S}'}}}\left( i,i' \right) \right):=v\left( \left( \mathcal{S}\backslash \left\{ i' \right\} \right)\cap \left\{ i \right\} \right)+ \\
    v\left( \left\{ i \right\} \right) -v({\mathcal{S}'})-v(\mathcal{S})
\end{multline}

\subsection{Algorithm Design}

Inspired by the Ref. \cite{ref11}, the distributed coalition algorithm (DCA) is designed as shown in Algorithm 1, which is run at each node. In lines 3 to 19 of the algorithm pseudocode, every node gathers potential positive gain operations that satisfy constraints and ultimately selects the trail operation with maximum operation gain. To determine whether to update the trail operation ${{{\mathsf{\hat{\mathcal{P}}}}}_{i,\mathcal{S}}}$ or maintain its original state, we introduce a log linear learning approach with probability $Pro\left( {{{\mathsf{\hat{\mathcal{P}}}}}_{i,\mathcal{S}}} \right)$ or updating and probability $1-Pro\left( {{{\mathsf{\hat{\mathcal{P}}}}}_{i,\mathcal{S}}} \right)$ for maintaining original operation ${{\mathsf{\mathcal{P}}}_{i,\mathcal{S}}}$. 

\begin{equation}
    Pro\left( {{{\mathsf{\hat{\mathcal{P}}}}}_{i,\mathcal{S}}} \right)=\frac{{{e}^{\frac{1}{\varepsilon }g\left( {{{\mathsf{\hat{\mathcal{P}}}}}_{i,\mathcal{S}}} \right)}}}{{{e}^{\frac{1}{\varepsilon }g\left( {{{\mathsf{\hat{\mathcal{P}}}}}_{i,\mathcal{S}}} \right)}}+{{e}^{\frac{1}{\varepsilon }g\left( {{\mathsf{\mathcal{P}}}_{i,\mathcal{S}}} \right)}}}
\end{equation}

The temperature $\varepsilon$ in (25) determines the amplitude of noise, indicating how likely a node is to take a sub-optimal action. A node would select the largest gain operation with arbitrarily high probability as $\varepsilon \to 0$. 

\begin{algorithm}[!t]  
\renewcommand\arraystretch{0.8}
	\caption{Distributed coalition algorithm}
	\LinesNumbered 
	\KwIn{current node $i$, current cluster $\mathcal{S}_{i}^{curr}$ for node $i$, last iteration operation ${{\mathsf{\mathcal{P}}}_{i,\mathcal{S}_{m}^{(i)}}}$, the number of clusters in current partition $M$, network graph $\mathsf{\mathcal{W}}$}
	  Find neighbor node set ${{\mathsf{\mathcal{R}}}_{i}}$ and neighbor cluster set $\mathsf{\mathcal{C}}_{i}^{neibr}=\left\{ \mathcal{S}_{i,l}^{neibr}\left| {{i}^{neibr}}\in \mathcal{S}_{i,l}^{neibr} \right.,\forall {{i}^{neibr}}\in {{\mathsf{\mathcal{R}}}_{i}} \right\}$\; 
   Initial the positive gain operation set ${{P}_{i}}=\varnothing $ \;
    \For{$\mathcal{S}_{i,l}^{neibr}\in \mathsf{\mathcal{C}}_{i}^{neibr}$}{ 
        \eIf{$v\left( \mathcal{S}_{i,l}^{neibr}\cap \left\{ i \right\} \right)>0$ and $g\left( {{\sigma }_{\mathcal{S}_{i}^{curr},\mathcal{S}_{i,l}^{neibr}}}\left( i,\varnothing  \right) \right)>0$}{
            //  the node $i$ join $\mathsf{\mathcal{S}}_{i,l}^{neibr}$ do not violate the constraint, and node $i$ obtain a positive gain by joining $\mathsf{\mathcal{S}}_{i,l}^{neibr}$ directly
            ${{P}_{i}}={{P}_{i}}\cup {{\sigma }_{\mathcal{S}_{i}^{curr},\mathcal{S}_{i,l}^{neibr}}}\left( i,\varnothing  \right)$
        }
        {
            \For{$j\in \mathcal{S}_{i,l}^{neibr}$}{
                \If{$v\left( \mathcal{S}_{i,l}^{neibr}\cap \left\{ j \right\} \right)>0$ and $g\left( {{\sigma }_{\mathcal{S}_{i}^{curr},\mathcal{S}_{i,l}^{neibr}}}\left( i,j \right) \right)>0$}{
                    //  switch operation 
                    ${{P}_{i}}={{P}_{i}}\cup {{\sigma }_{\mathcal{S}_{i}^{curr},\mathcal{S}_{i,l}^{neibr}}}\left( i,j \right)$ 
                }
                \If{$g\left( {{\rho }_{\mathcal{S}_{i}^{curr},\mathcal{S}_{i,l}^{neibr}}}\left( i,j \right) \right)>0$}{
                    //  replace operation
                    ${{P}_{i}}={{P}_{i}}\cup {{\rho }_{\mathcal{S}_{i}^{curr},\mathcal{S}_{i,l}^{neibr}}}\left( i,j \right)$
                }
            }
        }
    }
    \If{${{\varsigma }_\mathcal{S}}(i,i_\mathcal{S}^{h})>0$}{
        ${{P}_{i}}={{P}_{i}}\cup {{\varsigma }_\mathcal{S}}(i,i_\mathcal{S}^{h})$
    }
    Find  trail operation ${{\mathsf{\hat{\mathcal{P}}}}_{i,\mathcal{S}_{i}^{curr}}}$ obey $g\left( {{{\mathsf{\hat{\mathcal{P}}}}}_{i,\mathcal{S}_{i}^{curr}}} \right)>g\left( {{P}_{i,\mathcal{S}_{i}^{curr}}} \right)\quad \forall {{P}_{i,\mathcal{S}_{i}^{curr}}}\in {{P}_{i}}$\;
    Update coalition partition by operation ${{\mathsf{\hat{\mathcal{P}}}}_{i,\mathcal{S}_{i}^{curr}}}$ with probability in Eq. 23, otherwise maintain ${{\mathsf{\mathcal{P}}}_{i,\mathcal{S}_{m}^{(i)}}}$\;
    
\end{algorithm}

\subsection{Convergence analysis}

\textbf{Proposition 1.} \textit{For a VANET, starting from any coalition partition ${{\Pi }_{d}}$, the proposed distributed clustering algorithm based on coalition formation game always converges to a final coalition partition $\Pi $, which is composed of a number of disjoint coalitions.}

\textit{Proof: }Let ${{\Pi }_{t}}$ denote the coalition partition after $t$ iterations. Given the initial coalition partition ${{\Pi }_{d}}$, the proposed clustering algorithm is composed of a sequence of operations $\mathsf{\mathcal{P}}$. Therefore, the cluster formation phase consists of a sequence of operations such as the following:
\begin{equation}
    {{\Pi }_{d}}\to {{\Pi }_{1}}\to {{\Pi }_{2}}\to \cdots \to {{\Pi }_{t}}\to \cdots 
\end{equation}
Each switch operation results in an increase in the total coalition values of all coalitions. Considering that there is a finite number of partitions, it can be inferred that the sequence will always converge to a final coalition partition, which completes the proof.

The stability of the resulting coalition partition can be studied using the following stability notion from game theory \cite{ref6}. In other words, a partition is Nash-stable if there is no node has an incentive to deviate from its current coalition.

\textbf{Definition 5.} \textit{A partition $\Pi :=\left\{ {\mathcal{S}_{1}},...,{\mathcal{S}_{M}} \right\}$ a is Nash-stable if $\forall {\mathcal{S}_{m}}\in \Pi $, $\forall i\in {\mathcal{S}_{m}}$, $g\left( {{\mathsf{\mathcal{P}}}_{i,{\mathcal{S}_{m}}}} \right)\le 0$ for all ${{\mathsf{\mathcal{P}}}_{i,{\mathcal{S}_{m}}}}\in {{P}_{i}}$ }

\textbf{Proposition 2. } \textit{ Based on Proposition 1, the final coalition partition ${{\Pi }^{f}}$ is Nash-stable.}

\textit{Proof.} For the coalition partition $\Pi :=\left\{ {\mathcal{S}_{1}},...,{\mathcal{S}_{M}} \right\}$, no node $\forall i\in \mathsf{\mathcal{N}}$ has an operation incentive. Assume that the final coalition partition ${{\Pi }^{f}}$ from the proposed clustering algorithm is not Nash-stable. Then, there is an operation ${{\mathsf{\mathcal{P}}}_{i,{\mathcal{S}_{m}}}}\in {{P}_{i}}$ such that $g\left( {{\mathsf{\mathcal{P}}}_{i,{\mathcal{S}_{m}}}} \right)>0$. Hence, the node in coalition ${\mathcal{S}_{m}}$ can trigger an operation to improve the final partition, which contradicts the fact that ${{\Pi }^{f}}$ is the final partition. Therefore, the final coalition partition ${{\Pi }^{f}}$ resulting from Proposition 1 is Nash-stable.

\section{numerical experiments}
In this section, simulation setup of system model and algorithm parameter are firstly clarified. Then, experiments are carried out for the proposed distributed clustering algorithm (DCA) under different number of nodes in VANETs. The superiority of the DCA algorithm is then further demonstrated by comparing with the state-of-the-art algorithms. 

\subsection{Simulation Setup}
In this section, we use MATLAB for simulation to test the output of the proposed the proposed distributed clustering algorithm (DCA) compared to other existing coalition-game-based clustering algorithm. The experiment is simulated on a 5 km × 5 km square area where VANETs composed of different number of IVs and RSUs are distributed randomly and evenly. Table I lists the parameter values of simulation with reference to \cite{Wang.2020}. We adopt the distributed asynchronous model proposed in Ref. \cite{ref3}.

\begin{table}[]
\caption{Parameter Values}
\begin{tabular}{lll}
\hline
Symbol & Value & Specification                                         \\ \hline
$L$ & 5000$m$ & Length   of simulation area                           \\
$\left|   N \right|$      & 100$\sim$300   & Number   of nodes in the VANET                        \\
$Pr{{p}_{RSU}}$ & 0.1  & Proportion   of the number of RSUs                    \\
${{P}_{t}}$   & 30 dBm & Transmit   power                                      \\
$W$ & 800MHz & System   bandwidth                                    \\
${{N}_{0}}$   & -134dbm/MHz    & Background   noise                                    \\
${{\alpha   }_{L}}$ & 2 & Path   loss exponent                                  \\
$d_{ref}^{\min   }$, $d_{ref}^{\max }$ & 200$m$, 300$m$   & Minimal   distance and Maximal distance               \\
$v_{ref}^{\min   }$, $v_{ref}^{\max }$ & 20$m/s$, 30$m/s$ & Minimal   speed and Maximal speed                     \\
${{G}_{0}}$   & 20dBi & Maximum   antenna gain                                \\
$T$ & 500  & Number   of time slots                                \\
$\alpha   $   & 2 & Multi-hop   path loss parameter                       \\
$\beta   $  & 0.1  & Parameter   of inter-cluster link capacity            \\
${{V}_{intra}}$,   ${{V}_{inter}}$     & 1, 0.2 & \begin{tabular}[c]{@{}l@{}}Data   volume for inter-cluster and \\ intra-cluster management overhead\end{tabular} \\
${{N}_{max}}$,   ${{D}_{max}}$ & 15, 2 & \begin{tabular}[c]{@{}l@{}}Maximum   number of cluster member \\ and maximum cluster diameter\end{tabular}       \\ \hline
\end{tabular}
\end{table}

Existing algorithms compared with the proposed DCA are as follows:

1) Coalition selection algorithm based on coalition order (CSA-CO \cite{Ruan.2022}): according to the principle of first bilateral switch and then unilateral switch (randomly selected from exchangeable operations) to find the trial operation based on coalition order, and then every node in the network finally choose whether to accept the  according to log-linear algorithm. 
2) reliable and low-overhead clustering scheme (RLOC \cite{ref6}): With only unilateral switch, the most profitable operation is selected from the exchangeable operation to accept and update.
3) coalition selection algorithm based on coalition expected altruistic order (CSA-CEAO \cite{Chen.2021}): randomly form a partition as a initial clustering scheme, then employ only unilateral switch operation to explore all potential operation. Then, every node in networks update its action based on coalition expected altruistic order.
4) merge/split/transfer coalition formation algorithm (M/S/T-CFA \cite{Bassoy.2021}): consists of three different operations: Split, Merge, and Transfer. These three operations are repeated until convergence, which is reached within a finite number of iterations.

Among the aforementioned four comparison algorithms, CSA-CO, RLOC, CSA-CEAO, and our proposed algorithm are all distributed algorithm as they operate on every individual vehicle to obtain the clustering scheme. In contrast, M/S/T-CFA is a centralized algorithm that operates on a central processor to obtain the final clustering scheme.

\subsection{Simulation Results}
The statistical comparison results of baseline algorithm and the proposed DCA under different network scale are given in Table II, where \emph{best} and \emph{worst} represent the best and worst cases of the system objective respectively, $\bar{R}$ denotes the system objective average, and ${{\chi }_{total}}$ and ${{E}_{toal}}$ represent the average cooperative capacity value and average communication cost of the final clustering scheme out of 50 runs. 

The proposed DCA algorithm clearly performs best on all indicators and all scenarios and thus shows efficiency and robustness. Furthermore, as the number of vehicles in the network increases, the performance gap between the proposed algorithm and others also widens, highlighting its superiority in handling large-scale intelligent vehicle networks. Among all compared algorithms, only the centralized M/S/T-CFA algorithm comes close to matching the performance of PDCA. However, it should be noted that M/S/T-CFA requires traversing all potential merge/split/transfer operations of every node, resulting in significant computation time. Specifically, under scenario of 100 nodes, M/S/T-CFA takes 183.5 seconds to convergence while DCA completes within a mere 4 seconds.


\begin{table}[H]

\caption{Experimental results of algorithm comparison under different number of nodes without initial clustering}
\begin{tabular}{lllllll}
\hline
\begin{tabular}[c]{@{}l@{}}Node   \\    \\ num\end{tabular} & Algorithm & $\bar{R}$       & Best            & Worst           & ${{\chi }_{total}}$ & ${{E}_{toal}}$   \\ \hline
\multirow{5}{*}{100}                                        & DCA      & \textbf{0.9021} & \textbf{0.9240} & \textbf{0.8700} & \textbf{58.86}      & \textbf{980.00}  \\
 & RLOC      & 0.8443          & 0.8809          & 0.7921          & 53.34               & 1093.60          \\
 & CSA-CO    & 0.8559          & 0.8871          & 0.8177          & 54.48               & 1131.07          \\
 & CSA-CEAO  & 0.8001          & 0.8297          & 0.7662          & 49.13               & 1216.33          \\
 & M/S/T-CFA & 0.8789          & 0.9082          & 0.8457          & 56.64               & 1008.00          \\
\multirow{5}{*}{200}                                        & DCA      & \textbf{0.9143} & \textbf{0.9337} & \textbf{0.8917} & \textbf{138.15}     & \textbf{2627.53} \\
 & RLOC      & 0.8388          & 0.8722          & 0.8211          & 121.36              & 2911.47          \\
 & CSA-CO    & 0.8102          & 0.8384          & 0.7756          & 115.17              & 3469.73          \\
 & CSA-CEAO  & 0.7352          & 0.7764          & 0.6976          & 98.48               & 3688.13          \\
 & M/S/T-CFA & 0.8635          & 0.8732          & 0.8493          & 123.56              & 2753.57          \\
\multirow{5}{*}{300}                                        & DCA      & \textbf{0.9470} & \textbf{0.9732} & \textbf{0.8969} & \textbf{165.43}     & \textbf{5767.07} \\
 & RLOC      & 0.8816          & 0.9365          & 0.8484          & 148.95              & 6564.33          \\
 & CSA-CO    & 0.7865          & 0.8289          & 0.7472          & 125.05              & 8196.07          \\
 & CSA-CEAO  & 0.7124          & 0.7248          & 0.6851          & 106.24              & 8353.53          \\
 & M/S/T-CFA & 0.8931          & 0.9429          & 0.8413          & 152.94              & 6295.52          \\ \hline
\end{tabular}
\end{table}

We next focus on the convergence curves under 100 nodes of each algorithm. Fig. 4 shows that the average objective value of the DCA algorithm is better than that of the other algorithms after 10 time slots. Before 10 time slots, the DCA algorithm performs equally to RLOC, but better than CSA-CEAO and CSA-CO. CAS-CO has more randomness in the rules of updating agent actions (i.e. the log-linear learning rule), which leads to poorer performance in the early evolution process but a better final result relative to CAS-CO and RLOC. As a centralized algorithm, M/S/T-CFA is not included in Fig. 4 as its calculation iteration differs from that of other distributed algorithms under comparison as well as our proposed approach.

\begin{figure}[H]
	\centering
    \setlength{\abovecaptionskip}{0.cm}
    \includegraphics[scale=0.6]{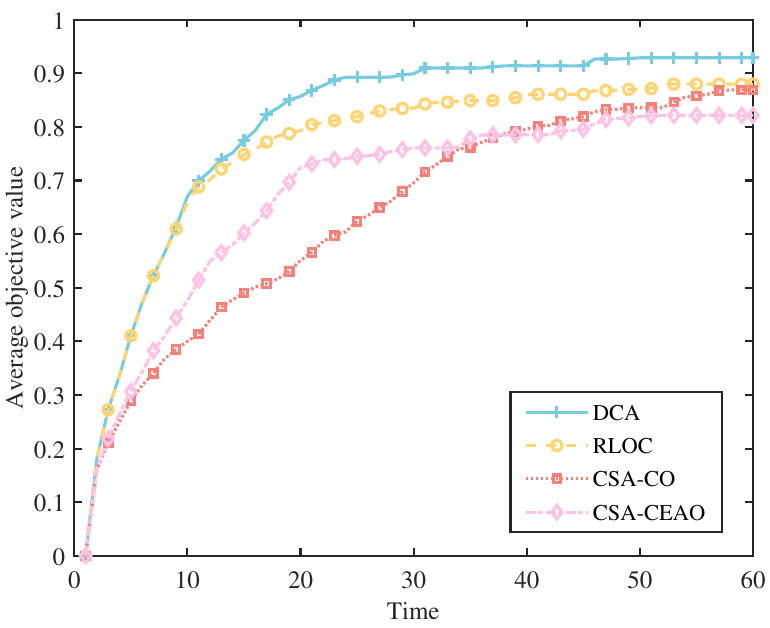}  
	\caption{The convergence curves under 100 nodes of each algorithm}   
\label{f101}
\end{figure}

The premise of the aforementioned experiments assumes that all algorithms lack an initial clustering scheme, implying that each node starts as an independent cluster. However, some comparative algorithms are executed with an input of an initial clustering scheme. To ensure fairness, we also compared the experimental outcomes of algorithms under the condition of having an initial clustering scheme. Consequently, the subsequent section will present a performance evaluation of different algorithms in a scenario with an initial clustering scheme using an initial clustering approach.

The initial clustering approach employs the clustering algorithm in the bootstrapping phase (CA-BP) in research \cite{Ergenc.2019}, which is based on the lowest-ID-based clustering algorithm initially proposed by Gerla et al. In CA-BP, each node broadcasts its unique node identifier (ID) and the node with the lowest ID is selected as a cluster head in each one-hop neighborhood after a quick convergence time. Besides, nodes residing in the coverage area of multiple clusters are designated as gateways and they become a member of the cluster with the lowest-ID cluster head.

From Table III, it is evident that different distributed clustering algorithms yield varying results after employing the CA-BP algorithm for initial clusters. In all cases, the proposed DCA algorithm outperforms the comparison algorithms with an initial clustering scheme generated by CA-BP. 
Comparing the results presented in Table 2 and Table 3 reveals that the proposed DCA algorithm achieves better objective values without relying on an initial clustering approach. This observation can be attributed to the unsatisfied initial clustering scheme generated by the CA-BP algorithm, which ultimately impacts the final clustering results of the DCA algorithm.

\begin{table}[]
\caption{Experimental results of algorithm comparison under different number of nodes with initial clustering}
\begin{tabular}{lllllll}
\hline
\begin{tabular}[c]{@{}l@{}}Node   \\    \\ num\end{tabular} & Algorithm & $\bar{R}$       & Best            & Worst           & ${{\chi }_{total}}$ & ${{E}_{toal}}$   \\ \hline
\multirow{5}{*}{100}                                        & DCA       & \textbf{0.8699} & \textbf{0.9168} & \textbf{0.8161} & \textbf{55.83}      & \textbf{1127.33} \\
 & RLOC      & 0.8007          & 0.8886          & 0.7427          & 49.21               & 1259.87          \\
 & CSA-CO    & 0.7878          & 0.8979          & 0.7270          & 46.86               & 1544.00          \\
 & CSA-CEAO  & 0.7529          & 0.7900          & 0.7340          & 43.50               & 1481.27          \\
 & M/S/T-CFA & 0.8129          & 0.8974          & 0.7949          & 50.21               & 1113.53          \\
\multirow{5}{*}{200}                                        & DCA       & \textbf{0.8558} & \textbf{0.8775} & \textbf{0.8162} & \textbf{125.38}     & \textbf{3512.53} \\
 & RLOC      & 0.7914          & 0.8495          & 0.7513          & 111.07              & 3775.00          \\
 & CSA-CO    & 0.7441          & 0.8299          & 0.7030          & 97.82               & 3776.53          \\
 & CSA-CEAO  & 0.7217          & 0.7701          & 0.6954          & 92.81               & 3719.00          \\
 & M/S/T-CFA & 0.8024          & 0.8465          & 0.7578          & 109.26              & 3758.63          \\
\multirow{5}{*}{300}                                        & DCA       & \textbf{0.8677} & \textbf{0.9063} & \textbf{0.8374} & \textbf{145.80}     & \textbf{8845.67} \\
 & RLOC      & 0.8164          & 0.8479          & 0.7728          & 133.04              & 10527.20         \\
 & CSA-CO    & 0.7380          & 0.7667          & 0.7112          & 113.96              & 13378.27         \\
 & CSA-CEAO  & 0.7219          & 0.7460          & 0.6971          & 106.01              & 14276.80         \\
 & M/S/T-CFA & 0.8187          & 0.8570          & 0.7753          & 129.74              & 10345.23         \\ \hline
\end{tabular}
\end{table}

\section{Conclusion}

We propose a distributed clustering algorithm (DCA) based on the coalition game theory for the clustering problem of intelligent vehicles based on the coalition formation game framework. We proved that the proposed algorithm converges to a Nash-stable solution while simultaneously satisfying some practical constraints, such as the cluster size, the cluster diameter, etc. In the case study and numerical experiments, we compared our algorithm with state-of-the-art coalition formation algorithms under different scales of VANET. The experiments showed that our algorithm was superior in terms of both the solution quality and the convergence efficiency.

\bibliographystyle{splncs04}
\bibliography{ref}

\end{document}